\documentclass[graybox]{svmult}

\usepackage{type1cm}        
\usepackage{makeidx}         
\usepackage{graphicx}        
\usepackage{float}           
\usepackage{multicol}        
\usepackage[bottom]{footmisc}

\usepackage{newtxtext}       %
\usepackage{newtxmath}       
\usepackage[utf8]{inputenc}

\usepackage{verbatim} 
\usepackage{lineno} 

\usepackage[square,numbers]{natbib}
\usepackage{url}

\usepackage[breaklinks]{hyperref}
\usepackage{breakurl}

\makeindex             

\begin{document}

\title*{The Role of Network Structure and Initial Group Norm Distributions in Norm Conflict}

\titlerunning{Network Structure, Group Norm Distribution, and Norm Conflict} 
\author{Julian Kohne, Natalie Gallagher, Zeynep Melis Kirgil, Rocco Paolillo, Lars Padmos, Fariba Karimi}
\authorrunning{J. Kohne, N. Gallagher, Z.M. Kirgil, R. Paolillo, L. Padmos, F. Karimi} 
\institute{Julian Kohne \at Department of CSS, GESIS, Leibniz Institute for the Social Sciences, Unter Sachsenhausen 6-8
50667 Cologne, Germany, \email{julian.kohne@gesis.org}
\and Natalie Gallagher \at Department of Psychology, Northwestern University, Evanston, IL 60208-2710, USA, \email{natalieg@u.northwestern.edu}
\and  Zeynep Melis Kirgil \at Department of Sociology, University of Groningen, Grote Rozenstraat 31 9712, TG Groningen, The Netherlands, \email{z.m.kirgil@rug.nl}
\and Rocco Paolillo \at BIGSSS, University of Bremen \& Jacobs University Bremen, UNICOM-Building, Haus 9, Mary-Somerville-Str. 9, 28359 Bremen, Germany, \email{rpaolillo@bigsss-bremen.de}
\and Lars Padmos \at Department of Sociology, University of Groningen, Grote Rozenstraat 31 9712, TG Groningen, The Netherlands, \email{l.padmos@student.rug.nl}
\and Fariba Karimi \at Department of CSS, GESIS, Leibniz Institute for the Social Sciences, Unter Sachsenhausen 6-8
50667 Cologne, Germany, \email{Fariba.Karimi@gesis.org}
\and Corresponding Author: Fariba Karimi (Forthcoming in Deutschmann, E., J. Lorenz, L.G. Nardin, D. Natalini and A.F.X. Wilhelm (eds). Computational Conflict Research. Cham: Springer Nature.)
}

%
%
\maketitle

\abstract*{Social norms can facilitate societal coexistence in groups by providing an implicitly shared set of expectations and behavioral guidelines. However, different social groups can hold different norms, and lacking an overarching normative consensus can lead to conflict within and between groups. In this chapter, we present an agent-based model that simulates the adoption of norms in two interacting groups. We explore this phenomenon while varying relative group sizes and homophily/heterophily (two features of network structure), and initial group norm distributions. Agents update their norm according to an adapted version of Granovetter's threshold model, using a uniform distribution of thresholds. We study the impact of network structure and initial norm distributions on the process of achieving normative consensus and the resulting potential for intragroup and intergroup conflict. Our results show that norm change is most likely when norms are strongly tied to group membership. Groups end up with the most similar norm distributions when networks are heterophilic, with small to middling minority groups. High homophilic networks show high potential intergroup conflict and low potential intragroup conflict, while the opposite pattern emerges for high heterophilic networks.}

\abstract{Social norms can facilitate societal coexistence in groups by providing an implicitly shared set of expectations and behavioral guidelines. However, different social groups can hold different norms, and lacking an overarching normative consensus can lead to conflict within and between groups. In this chapter, we present an agent-based model that simulates the adoption of norms in two interacting groups. We explore this phenomenon while varying relative group sizes and homophily/heterophily (two features of network structure), and initial group norm distributions. Agents update their norm according to an adapted version of Granovetter's threshold model, using a uniform distribution of thresholds. We study the impact of network structure and initial norm distributions on the process of achieving normative consensus and the resulting potential for intragroup and intergroup conflict. Our results show that norm change is most likely when norms are strongly tied to group membership. Groups end up with the most similar norm distributions when networks are heterophilic, with small to middling minority groups. High homophilic networks show high potential intergroup conflict and low potential intragroup conflict, while the opposite pattern emerges for high heterophilic networks.}


\section{Introduction}
\label{sec:summary}

In this chapter, we study the impact of network structure and initial group norm distributions on the process of arriving at a normative consensus between groups and the potential for intragroup and intergroup conflict that might emerge under different conditions. To this end, we first provide a brief theoretical overview on social norms, normative group conflict and the process of finding consensus trough social influence. Secondly, we give an overview on the role that network structure as well as the initial distributions of norms can play in this process. Specifically, we argue that homophily/heterophily (preference for forming connections to similar/dissimilar others) between members of different groups, relative group sizes, and the initial distribution of norms within groups are all important factors for reaching normative consensus, and consequently relevant determinants of conflict potential. Based on this reasoning, we develop an agent-based model that simulates social networks of agents from two different social groups where each agent holds one of two social norms. In an adapted version of Granovetter's threshold model \cite{granovetter1978threshold}, each agent updates its social norm by comparing the proportion of norms held by its immediate neighbors to an internal threshold drawn from a uniform distribution. Agents are thus "observing" the "openly displayed behavior" of their neighbors and adapt their own behavior accordingly if enough of their neighbors display a different norm. We apply this model to different network structures, defined by relative group sizes and homophily/heterophily between agents from different groups. This allows us to assess the impact of these structural network properties on the process of reaching normative consensus and associated conflict potential. In addition, we run our model for different levels of initial group norm distributions, so that we can also assess the influence of alignment (or independence) of norms and social group membership. We define and examine three relevant outcomes: The degree to which norm distributions change, the degree to which the difference in norm distributions between the two groups changes, and the potential for conflict within and between the groups. Lastly, we discuss our results with respect to their applicability, the limitations of our model and possible directions for future research.

\section{Social Norms}
\label{sec:social_norms}

\textit{Social norms} can be defined as unwritten behavioral rules \cite{bicchieri2014norms} or "social standards that are accepted by a substantial proportion of the group” \cite{forsyth2018group}. They are a shared set of situation-specific behaviors that facilitate social interaction by providing an implicitly shared set of expectations and behavioral guidelines \cite{bicchieri2005grammar}. Such behaviors can range from an implicit dress code at work, to the expression of religious and political symbols, or (not) interacting with other social groups. Norms are implicitly negotiated between members of a group and enforced through informal sanctions, such as gossip, censoring or ostracism \cite{bicchieri2005grammar}. They are passed through generations via socialization processes in childhood \cite{house2018social} and are, in contrast to laws, not necessarily enforced by an institution. Norms come in multiple types; for example, \textit{prescriptive norms} define behaviors that one should enact (e.g. "offering elderly people a seat on the subway"), while \textit{proscriptive norms} define undesirable behaviors that one should avoid (e.g. "interrupting people while they speak"). The most important distinction for our purposes is between \textit{injunctive} and \textit{descriptive} norms. Injunctive norms focus on beliefs about how people should act, while descriptive norms are defined by the observation of how people actually do act \cite{melnyk2010influence,cialdini1990focus}. For instance, "everybody should recycle" is an injunctive norm, while the observation that many people do not recycle represents a descriptive norm \cite{cialdini1990focus}. Both types of norms are important determinants of behavior, but previous research suggests that injunctive norms primarily elicit behavioral change by changing attitudes \cite{melnyk2010influence,megens2010attitudes}, while descriptive norms directly impact behavior \cite{cialdini2007influence}. In this chapter, we are interested in descriptive norms, because they are directly inferred from the observed behavior of others. Injunctive norms can differ from directly observed behavior, and can involve more complex cognitive processes \cite{house2018social}, which are beyond the scope of our model. Therefore, when we are referring to social norms with respect to our model, we are specifically addressing descriptive social norms.

\subsection{Normative Conflict}
\label{sec:normative_conflict}

A large body of previous research has focused on the potential for positive impact of social norms on behavior. Predominantly, these studies were interested in changing individual beliefs or behavior by presenting normative information at odds with the individual's current beliefs or behavior. Examples include the reinforcement of non-delinquent behavior through the influence of peers \cite{megens2010attitudes}, positive effects of punishment on cooperative behavior \cite{fehr2000cooperation}, effects of social norms on compliance to vaccination programs \cite{OrabyThampiBauch2014}, reduction of binge-drinking in college students \cite{haines1996changing} and littering \cite{cialdini1990focus}. However, inconsistent norms do not only elicit behavioral change; they can lead to interpersonal and intergroup conflict\cite{hogg2006social}. The potential risk of such normative conflicts is especially high in multicultural contexts where different cultural groups must coexist \cite{wimmer2013ethnic}. A recent example of normative conflict in Europe is women wearing a veil to cover their face in public. This practice is a prescriptive social norm in some predominantly Muslim countries and it has elicited mixed reactions when immigrants engaged in the practice in their new countries \cite{kilicc2008introduction}. Some western countries such as France, Belgium and Switzerland have banned the practice. In France, lawmakers claimed that a ban was necessary to ensure \textit{"peaceful cohabitation"} \cite{zeit_burqa}. Likewise, in Germany, face veils have been controversially discussed in the past years: For instance, the German Minister of the Interior stated \textit{"[...] we reject this. Not just the headscarf, any full-face veils that only shows eyes of a person [...] It does not fit into our society for us, for our communication, for our cohesion in the society ... This is why we demand you show your face"} \cite{cnn_veil}. This backlash reflects an underlying normative conflict, with a large majority (81\%) of Germans supporting a ban in public institutions and a substantial group (51\%) even supporting a general ban. Only a minority of the national population (15\%) indicate that they are not in favor of any kind of regulation \cite{dimap_burqa}.

However, such normative societal conflicts exist not only along established cultural and religious divides, but can cover a wide array of topics and elicit intergroup and intragroup conflicts \cite{hogg2006social}. For instance, gun ownership is a controversial normative debate within U.S. society \cite{kleck1996crime}, involving subgroups with different cultural orientations \cite{celinska2007individualism}. Abortion is another topic debated worldwide, with disagreements concerning womens rights, health care systems and moral constraints \cite{marecek2017abortion}. Empirical research shows how the controversy around abortion leads to a polarization of opinions within Protestants and Catholic groups in U.S. society \cite{evans2002polarization}. Other inconsistent norms can concern controversial national traditions such as \textit{Zwarte Piet} ("Black Pete"), a folklorist character and helper of \textit{Sinterklaas} (Santa Claus) in the Dutch culture. The character is typically displayed with blackface makeup, bright red lips and colorful clothing. The display has been increasingly criticized as a racist stereotype, predominantly by minority and immigrant groups, while many native Dutch citizens argue that "Black Pete" is a positive character and part of their national tradition \cite{rodenberg2016essentializing}. In essence, inconsistent social norms within a larger collective have the potential to lead to intergroup, as well as intragroup conflict. With respect to trends of increasing globalization and migration, effectively resolving these normative conflicts is becoming a striking priority for many societies in the future. 

\subsection{Finding Consensus}
\label{sec:finding_consensus}

Despite their potential for negative outcomes, normative conflicts are not an indication that a collective is inherently unfit to live together peacefully. In contrast, they can be fundamental to the formation of social units at different scales. Georg Simmel defines shared consensus on social roles and their supporting norms as necessary features of human society \cite{simmel2009sociology}. Similarly, normative conflicts are frequently observed in the literature on group formation and described as a necessary step towards a common group identity. For example, in Tuckman's stage model of group development, the \textit{norming stage} focuses on resolving disagreement and establishing a shared set of behavioral guidelines; it is a crucial step in the formation of an effective group \cite{tuckman1965developmental}. Some recent, empirically validated models such as the  Normative Conflict Model \cite{packer2014tough} confirm this mechanism. According to the model, members strongly identified with the group are more likely to openly express dissent compared to weakly identified members \cite{packer2014tough}. Dissenters help uncover the causes of the conflict and discuss possible solutions. To form an effective group with committed members, it is necessary to effectively resolve conflicts due to incompatible norms and to find a consensus on which most members agree. Failure to reach such a consensus might result in a lack of common group identity and task effectiveness, leading to the dissolution of the group \cite{tuckman1965developmental}.

Interactions between people from different social groups are a steadily increasing occurrence in societies that are socially, economically and culturally diverse \cite{arapoglou2012diversity}. Such diversity is likely to increase in the future, along with changing relations between majority and minority groups due to demographic and socioeconomic changes \cite{crul2016super}. As ongoing political and societal polarization in Western societies already demonstrate, incompatible social norms associated with different groups have the potential to elicit conflict \cite{fiorina2008political}. For these reasons, we argue that it is crucial to understand the conditions enabling social groups to effectively reach a normative consensus and how this process relates to conflict potential within and between social groups.

\section{Network Structure \& Group Norm Distributions}
\label{sec:network_structure}

Individuals do not adopt norms in isolation; the structure of their social environment is a key determinant of social behavior. The social networks in which we are embedded determine the kinds of people and behavior to which we are exposed, thereby shaping the descriptive norms we hold. Thus, the interpersonal processes which contribute to finding normative consensus \cite{neumann2008homo}, as well as the intergroup and intragroup processes \cite{hogg2006social}, are crucially contextualized within networks of social interaction. Consequently, we argue that finding normative consensus is a continuous process of group members mutually exerting social influence \cite{cialdini2007influence} on each other until a relatively stable equilibrium is reached \cite{latane1981psychology,flache2017models}. This often requires that at least some individuals react to social influence exerted on them by their social networks by changing their norms. For instance, \cite{kalesan2016gun} show how networks such as family and friends are the best predictors in forging a culture favoring gun ownership. As for the normative conflict of gay marriage in the U.S., a longitudinal time-series study shows how the decision of the U.S. Supreme Court in June 2015 eventually led to an increase in perceived social norms supporting gay marriage independently of individual attitudes \cite{TankardPaluck2017}. In short, the social networks people are embedded in appear to play a crucial role in the process of reaching a normative consensus within and between groups.

In this chapter, we will focus on homophily/heterophily between people from different groups and relative group sizes as determinants of network structure, and on the initial distribution of norms within groups when they come into contact.

\subsection{Homophily \& Heterophily}
\label{sec:homophily_heterophily}

Homophily is the tendency to preferentially connect and interact with similar others \cite{mcpherson2001birds}, while heterophily is the tendency to preferentially connect and interact with dissimilar others \cite{lozares2014homophily}. Homophily has been observed extensively in many social networks, including school friendships \cite{stehle2013gender}, scientific collaborations \cite{jadidi2017gender}, and online communications \cite{mislove2010you}. It is likely a manifestation of the \textit{similarity bias}, a fundamental human tendency to like and value others that are similar to the self and to consequently be disproportionally influenced by them \cite{cialdini2007influence}. For example, a controlled experimental study on the spread of a health innovation through social networks varied the level of homophily, showing that homophily significantly increased the overall adoption of new health behavior, especially among those in more clustered networks \cite{centola2010spread}. Similar effects have been shown in diverse health behaviors in large social networks, such as the spread of smoking \cite{christakis2008collective} and obesity \cite{christakis2007spread}. Since social influence is exerted through social ties in networks \cite{aral2011creating,lewis2012social} and homphily/heterophily determines how these ties are formed, we argue that it is an important factor in the process of negotiating a normative consensus through mutual social influence.

\subsection{Group Size}
\label{sec:group_size}

Almost no collective group is made out of completely homogeneous members. Instead, they consist of demographic subgroups, such as those defined by gender, nationality, or education \cite{mcpherson2001birds}. Mostly, these subgroups are not equally sized, so that people are either part of a majority or minority group \cite{blau1977macrosociological} with respect to a certain social category. The pervasive influence of majority opinions, customs, and norms is well established in theoretical accounts of group-based social influence \cite{latane1981psychology}. The dominant role of the majority has been experimentally validated in numerous studies replicating the seminal work by \cite{asch1951effects}, both for individual social influence  \cite{horcajo2010effects,kundu2013morality} and group influence \cite{meyers2000majority,cohen2003party}. Greater influence of the majority is generally assumed for acculturation processes of minority immigrants in host countries \cite{bourhis1997towards,ward2010contextual}. Yet, other studies have demonstrated that under certain conditions, minorities can successfully exert social influence on the majority and consequently redefine the normative consensus in their favor \cite{hogg2006social, mugny1982minority, nemeth1986differential}. For these reasons, we argue that the sizes of interacting subgroups within a larger society are an important factor in the process of negotiating normative consensus.

\subsection{Initial Group Norm Distributions}
\label{sec:norm_endorsement_proportions}

Agreement on social norms is considered to be a part of the collective identity people derive from the social groups to which they belong \cite{hornsey2008social,hogg2006social}. Norms vary, however, in how much they align with group membership. Even in the case of German opinions on face veils, a full 15\% do not agree with the normative opinion to ban face veils \cite{cnn_veil, dimap_burqa}. That is, despite sharing group membership, individuals disagree on this norm.  Conversely, in the social group of Muslim immigrants in Germany, some will support the norm of face veils while others will oppose it. People can hold the same norm on face veils even though they are from different social groups, or they can hold different social norms while belonging to the same social group. In terms of our example, there will be some Muslim immigrants agreeing with Germans who oppose face veils. There will also be some Germans agreeing with the Muslim immigrants who do not oppose face veils. In short, even in this case of strong consensus, group membership is not the single determinant of norms held on an individual level. Social norms are often aligned with group membership to a degree, but the two are not synonymous. 

This interplay of social group membership versus agreement in moral or normative issues has been shown to be influential in previous studies. For instance, the influence that a group exerts on individuals is not only a function of its size, but also of its unanimity, with stronger pressure towards conformity for more unanimous groups \cite{asch1956studies}. Furthermore, studies have shown that people react more negatively to dissenters from their own in-group \cite{marques1988black} and consequently punish them harder. The initial distribution of norms within groups thus seems to be important for negotiating a normative consensus, even though it is not necessarily influencing the structure of the social network.

\section{Agent-Based Model} 
\label{sec:simulating_norm-based_conflict}

Agent-based modeling can be of particular interest to understand social phenomena because it enables researchers to study complex macro-level outcomes that emerge from a clearly defined set of micro-level processes \cite{macy2002factors,flache2017models}. In addition, simulations allow us to systematically vary agents' behavioral rules or the circumstances in which they act \cite{squazzoni2014social}. In short, agent-based models help us to gain insight into the emergence of complex systems by systematically testing a variety of different parameters and the combined impact they exert on the emergent system \cite{macy2002factors}. Previous research has extensively used agent-based models to study phenomena such as spatial segregation \cite{schelling1971dynamic}, opinion diffusion \cite{lorenz2007continuous}, the adoption of innovation \cite{zhang2017empirically}, and cascade effects \cite{watts2002simple}.

For the purpose of modeling normative conflict in social networks with respect to relative group sizes, homophily/heterophily, and group norm differences, we developed a modular simulation framework based on a network generation algorithm using preferential attachment, group size and homophily/heterophily \cite{karimi2018homophily}, and Granovetter's threshold model \cite{granovetter1978threshold}. We utilized R \cite{Rlang} for our model as it appears to be more widespread among the social science community than Python and offers more customizability, better parallelization and scalability than NetLogo. Consequently, probabilistic processes in our model are implemented using the \textit{sample()} function in R, which relies on the current system time to generate a seed for pseudo-random number generation. All code, documentation and an animated visualization are available on GitHub \cite{SimGit} under the MIT License.

\subsection{Simulating Norm Conflict}
\label{sec:agent-based_model}

In our agent-based model, we aim to simulate the impact of group size, homophily/heterophily between agents from different groups, and initial group norm distributions on the process of reaching normative consensus and resulting conflict potential. To this end, we generated networks with 2000 agents each, where network structure is determined by one parameter for relative group size ($g$) and one parameter for homophilic/heterophilic preferences of agents ($h$) \cite{karimi2018homophily}. In addition, initial norms for agents were assigned based on three different pairs of binomial probabilities, resulting in three conditions for initial group norm distributions. Once the network structure is generated and agents are assigned their initial norms, each agent is assigned a threshold from a uniform distribution \cite{granovetter1978threshold} and the model simulates normative social influence processes between agents by repeating 50 iterations of Granovetter's threshold model. Once the simulation is complete, we extract the percentage of agents holding each norm for each group, and the number of ties between agents within each group and between the groups. Crucially, we differentiate ties between agents holding the same norm and ties between agents with incompatible norms. Our model thus consists of four subsequent steps: Generation of network structure, initialization of group norm distributions, the norm updating process, and the extraction of outcome metrics.

In total, we simulate 150 unique parameter combinations with 20 networks per combination, resulting in 3000 unique networks (for an overview of the parameter space, see Table \ref{table:Parameters}). For each of these networks, we are saving each iteration of Granovetter's threshold model as an individual network object, resulting in 150.000 networks with 2000 agents each. Simulation was carried out on the High Performance Computing Cluster of the University of Cologne on 150 MPI nodes. We opted for 50 iterations of Granovetter's threshold model because it was the highest number of feasible iterations in the maximum computation time limit for the MPI nodes (360 hours) of the High Performance Computing Cluster. The simulation took approximately 13 days (315 hours) and resulted in approximately 40GB of output data.

\begin{table}[!t]
\caption{Range of parameter values of the simulation in the experiment}
\label{table:Parameters}
\begin{tabular}{p{2cm}p{5.2cm}p{4cm}}
\hline\noalign{\smallskip}
 Parameter & Description & Value(s) \\
 \noalign{\smallskip}\svhline\noalign{\smallskip}
 n  & No. of Agents in Network & 2000  \\
 m & Minimum Agent Degree & 2 \\
 p\textsubscript{1}:p\textsubscript{2}$^a$ & Initial Group Norm Distribution & [0.5:0.5][0.6:0.4][0.8:0.2] \\
 t & Individual Agent Threshold & $U(0,1)$ $^b$ \\
 g & Group Size &  [$0.1,\stackrel{+0.1}{\dots},0.5$] \\
 h & Homophily/Heterophily Parameter & [$0.1,\stackrel{+0.1}{\dots},1$] \\
 \noalign{\smallskip}\hline\noalign{\smallskip}
    \end{tabular}
$^a$ Each of the three conditions compares different initial distribution of the majority norm in the majority group (p\textsubscript{1}) and in the minority group (p\textsubscript{2})\\
$^b$ $U$:Uniform distribution
\end{table}

\subsection{Generation of Network Structure}
\label{sec:generating_network_structure}

To generate different network structures that resemble real social networks and enable comparison of effects of $g$ and $h$, we implemented the network generation algorithm by \cite{karimi2018homophily}. This algorithm combines the \textit{preferential attachment} mechanism, which has been observed in many large-scale social networks \cite{barabasi1999emergence}, with tunable parameters for group sizes and homophilic/heterophilic tendencies of agents in the model. As a point of terminology, we will refer to the group containing more agents as the "majority group" and the group containing less agents as the "minority group".

The network generation model implements an iterative growth process where we start out with a small number of $m$ initial agents for both the majority group and the minority group. After this initial setting, one agent is added to the network at a time. Each new agent has a probability of $g$ to be assigned to the minority group and a probability of $1 - g$ to be assigned to majority group. For example, with a value of $g = 0.4$, each new agent has a probability of 40\% to be assigned to the minority group and a probability of 60\% to be assigned to the majority group. Each new agent forms $m$ ties to the agents that are already present from previous steps. In this way, the parameter $m$ also defines the \textit{minimum degree} of agents in the network. We keep this parameter constant at $m=2$ across all our generated networks because it ensures that no agent is isolated in the network. Previous research demonstrated that the choice of $m$ does not change the properties of the network \cite{barabasi1999emergence}.

Connecting these $m$ ties from the new agent to existing agents is probabilistic, and relies on the homophily parameter $h$ and the degree of the present agents \cite{karimi2018homophily}. The parameter $h$ ranges from 0 to 1 and defines the likelihood of agents to form ties to agents from the same group or from a different group ($1 - h$). A value of 0 represents perfect heterophily (ties will only be formed between agents assigned to different groups) and a value of 1 represents perfect homophily (ties will only be formed between agents assigned to the same group). In addition, agents also have a build-in preference for agents with high degree (preferential attachment), which is interacting with their group preference determined by $h$. Specifically, the probability  $p_{ij}$ of each added agent $j$ to form a tie with a present agents $i$, depends on the degree of the present agent ($k_i$) and the specified homophily parameter between $i$ and $j$, $h_{ij}$, divided by the sum over all existing agents denoted by ($l$):

\begin{equation}
p_{ij} =
  \frac{h_{ij}k_i}
  {\sum{h_{lj} k_l}}\;.
\end{equation}

\begin{figure}[H]
    \centering
    \includegraphics[angle=90,width=\textwidth]{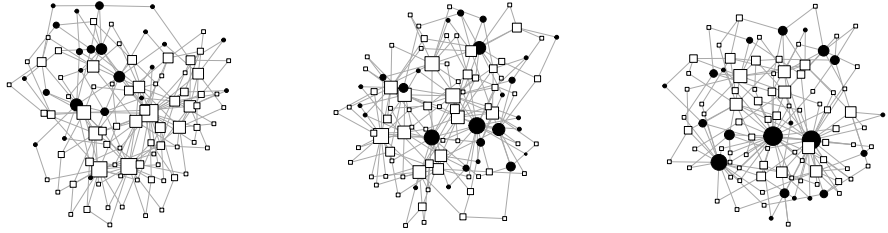}
	\caption{Generated networks with 100 agents and $g = 0.2$. From left to right, the networks are showcasing $h = 0.2$, $h = 0.5$ and $h = 0.8$. Node size represents logarithmized agent degree. Minority group agents (20\%) are represented by black circles, majority group agents (80\%) are represented by  white squares. When the network is heterophilic (left), the minority group increases their degree rapidly due to the combination of preferential attachment and smaller group size. In the homophilic network (right), the minority group cannot grow their degree by attracting majority group agents} 
	\label{fig:Heffect}
\end{figure}

The processes of assigning agents to a group and selecting present agents to connect with are not deterministic, so the same set of initial parameters will generate slightly different network structures each time. To capture this variance, we generate 20 networks per parameter combination and report averaged results. See Figure \ref{fig:Heffect} for an example, and see appendix for analytical derivations.

\subsection{Initialization of Group Norm Distributions}
\label{sec:initialization_of_group_devisiveness}

After creating network structures based on the parameters $g$ and $h$, we initialize a norm as an attribute in each agent. We will use "majority norm" and "minority norm" when we discuss our results with respect to the two different norms in our model. Specifically, majority norm will refer to the norm held by the larger proportion of agents in the larger of the two groups after initializing the network structure. In cases where the amount of agents holding each norm is equal, we simply track one of the two norms over the course of the simulation.

We use a probabilistic process with two different parameters $p\textsubscript{1}$ and $p\textsubscript{2}$ for the initial group norm distributions, where $p\textsubscript{1}$ describes the probability of agents in the majority group to be assigned the majority norm, while $1 - p\textsubscript{1}$ describes the probability of agents in the majority group to be assigned to the minority norm. Vice versa, $p\textsubscript{2}$ describes the probability of agents in the minority to be assigned the majority norm while $1 - p\textsubscript{2}$ describes the probability of agents in the minority being assigned the minority norm. For example, with $p\textsubscript{1} = 0.7$ and $p\textsubscript{2} = 0.3$, each agent in the majority group has 70\% probability of being assigned the majority norm and probability 30\% of being assigned the minority norm. Conversely, each new agent assigned to the minority has a probability of 30\% to be assigned the majority norm and probability of 70\% to be assigned the minority norm. In this example, we can see that $p\textsubscript{1}$ and $p\textsubscript{2}$ define how closely the assignment of norms is related to the group membership of new agents. If $p\textsubscript{1}$ and $p\textsubscript{2}$ are both 0.5, then there is no connection between group membership and norm - every agent of either group has an equal probability (50\%) to endorse either norm. If $p\textsubscript{1}$ is large and $p\textsubscript{2}$ is small, then initial norm proportions are associated with group membership - the majority and the minority group preferentially use different norms. In our model, we will be testing one case where the initial norm distribution is unrelated to group membership ($p\textsubscript{1} = 0.5$ and $p\textsubscript{2} = 0.5$), one where the initial norm distribution is weakly related to group membership ($p\textsubscript{1} = 0.6$ and $p\textsubscript{2} = 0.4$) and one where initial norm distribution is strongly related to group membership ($p\textsubscript{1} = 0.8$ and $p\textsubscript{2} = 0.2$). We thus generate models where (a) 50\% of the majority group and 50\% of the minority group start with the majority norm, (b) 60\% of the majority group and 40\% of the minority group start with the majority norm, and (c) 80\% of the majority group and 20\% of the minority group start with the majority norm.

\subsection{Norm Updating Process}
\label{sec:norm_updating_process}

After initializing one of the two norms in each agent according to parameters $p\textsubscript{1}$ and $p\textsubscript{2}$, we simulate the adoption of norms over time within each network using Granovetter's threshold model \cite{lewis2012social,aral2011creating,dimaggio2012network}. In our simulation, we use a modified version \cite{granovetter1978threshold} where each agent in the model is assigned a threshold value from a uniform distribution [0,1]. A central point in Granovetter's threshold model is the \textit{variability} of thresholds within a group. Once people with lower thresholds adopt a norm, they will raise the proportion of people with that norm, increasing the chance of shifting those who have higher thresholds \cite{granovetter1978threshold}. In his seminal work, \cite{granovetter1978threshold} showed these dynamics both with a uniform distribution and a normal distribution of thresholds. In our model, we decided to use a uniform distribution of thresholds because our aim is to understand the role of network structure and initial norm distributions in normative conflict, and not primarily to investigate the effects of the threshold. To clearly understand emergent properties in agent-based models without extraneous mechanisms, it is beneficial to avoid unnecessary complexities \cite{erleben2017simulate}. Non-uniform distributions require particular choices: either the single value of the threshold held by all agents, or the mean and variance of a normally-distributed threshold parameter. Thus, any distribution besides the uniform requires additional assumptions without adding a concrete contribution \cite{railsback2019agent} to our research questions. We use a uniform distribution in our model to control the effect of the threshold distribution \cite{lee2015complexities} while testing the effect of network structure and initial norm distribution. We also allow agents to change back and forth between norms as appropriate given their threshold and the norms of their neighbors. This is distinct from some models where an agent can only change once (e.g. learning of a new innovation), and we consider it appropriate for modeling our phenomenon of interest - descriptive social norms.

In the updating process, each agent compares its threshold value to the proportion of its immediate neighbors holding a particular norm. If the proportion of neighbors that are expressing a given norm is equal to or higher than the agent's threshold, the agent will update its currently held norm. For example, if agent $j$ has a threshold of $t\textsubscript{j} = 0.6$, it will update to the norm that 60\% or more of its neighbors display. Depending on the current norm of the agent, this can either mean switching to a different norm or keeping the agent's current norm. If both proportions fail to reach the threshold (e.g. 50/50 distribution of norms in neighborhood of agent while the threshold value is $0.6$), the agent will also keep the current norm. In cases where observed proportions of both norms are equal and exceeding an agents threshold, the agent will choose one of the two norms at random. Each network goes through $50$ iterations of the updating process, so all agents update their norms 50 times.

In each iteration of the norm updating process, all agents are updated \textit{asynchronously}, meaning that only one agent is updated at a time and the order in which agents update their norms is randomly shuffled before each iteration of the updating process. Thus, each agent's updating process can affect the updating process of the next agent. We chose this procedure as opposed to having a fixed order for updating agents or updating all agents at the same time because natural social interactions neither occur in a predetermined order nor do all people in a social network exert influence on each other simultaneously. For this reason, we argue that our approach more closely resembles real-life interactions and social influence processes between people.

\subsection{Outcome Metrics}
\label{sec:outcome_metrics}

After the agent-based model finishes, we extract our outcomes of interest: The degree to which norm distributions change, the degree to which the difference in norm distributions between the two groups changes, and the potential for conflict within and between the groups.

To operationalize the degree to which norm distributions change, the initial proportion of agents holding the majority norm is subtracted from the final proportion of agents holding the majority norm. We subsequently call this \textit{Change in Majority Norm} because it expresses the degree to which the group has adopted the majority norm relative to the group's starting point. If this number is positive, the group's use of the majority norm has increased over the course of the simulation. For example, if the network starts with an 80-20 group norm distribution and ends with 60\% of the minority group endorsing the majority norm, the minority group has adopted the majority norm by 40\%. If change in majority norm is negative, the group has rejected the majority norm. In a similar example, if the network starts with an 80-20 group norm distribution and ends with 10\% of the minority group endorsing the majority norm, the minority group has rejected the majority norm by 10\%. This is a group-level outcome: it tells us how the normative consensus within the majority group and within the minority group have changed over time. It is worth noting that the initial norm distribution limits the possible change within the majority group and the minority group. In the 80-20 initial norm distribution, only 20\% more of the majority group could hold the majority norm, while 80\% more of the minority group could do so.

At a system level, we are interested in the degree to which the difference in norm distributions between the two groups changes. Specifically, we are interested in whether the two groups express the two norms in similar proportions after the last iteration and if they have become more similar in their norm proportions over time. To calculate this, we first calculate the initial group norm difference by subtracting the initial proportion of the minority group holding the majority norm from the initial proportion of the majority group holding the majority norm $\Delta(p)_{\textsubscript{initial}} = p_{1_{\textsubscript{initial}}} - p_{2_{\textsubscript{initial}}}$ (see section \ref{sec:initialization_of_group_devisiveness}). Then we calculate the final group norm difference by subtracting the final proportion of the minority group holding the majority norm from the final proportion of the majority group holding the majority norm, $\Delta(p)_{\textsubscript{final}} = p_{1_{\textsubscript{final}}} - p_{2_{\textsubscript{final}}}$. We subtract the final group norm difference from the initial group norm difference to define \textit{Change in Group Norm Difference} $\Delta(p)_{\textsubscript{final}} - \Delta(p)_{\textsubscript{initial}}$. If this is positive, then difference has increased; the groups have become less similar over the course of the simulation in terms of their norms. If this is negative, then the group norm difference has decreased; the groups have become more similar. Once again, it is worth noting that the initial group norm distribution limits total possible change. 

At a dyadic level, we are interested in the potential for interpersonal conflict between and within groups. To look at this, we define \textit{Conflict Ties} as ties connecting two agents with different norms after the last iteration. Crucially, we distinguish between conflict ties of agents from the same group as a proxy for potential intragroup conflict, and conflict ties of agents from different groups as a proxy for potential intergroup conflict. In particular, we are extracting the proportion of ties in the majority group that connect agents with inconsistent norms, the proportion of ties in the minority group that connect agents with inconsistent norms, and the proportion of ties between the groups that connect agents with inconsistent norms.

\section{Simulation Results}
\label{sec:simulation_results}
Our results are structured around the three overarching outcome metrics outlined above. For each metric, we consider the aggregated output of our runs by averaging over the values obtained from the 20 simulated networks per parameters combination.

\begin{enumerate}
\item{\textbf{Change in Majority Norm:}}
Which combinations of parameters increase or decrease the prevalence of the majority norm? In which cases does the majority norm become prevalent among the majority and the minority group? In which cases does the minority norm gain prevalence?
\item{\textbf{Change in Group Norm Difference:}}
Which combinations of parameters reduce between-group norm differences? Which make convergence of norms most likely?
\item{\textbf{Conflict Ties:}}
Which sets of parameters make it most likely that potential within-group or between-group conflict will emerge? Which make it most likely that there will be little potential for conflict?
\end{enumerate}

\subsection{Change in Majority Norm}
\label{sec:change_majority_norm}

\begin{figure}
    \centering
    \includegraphics[width=12cm]{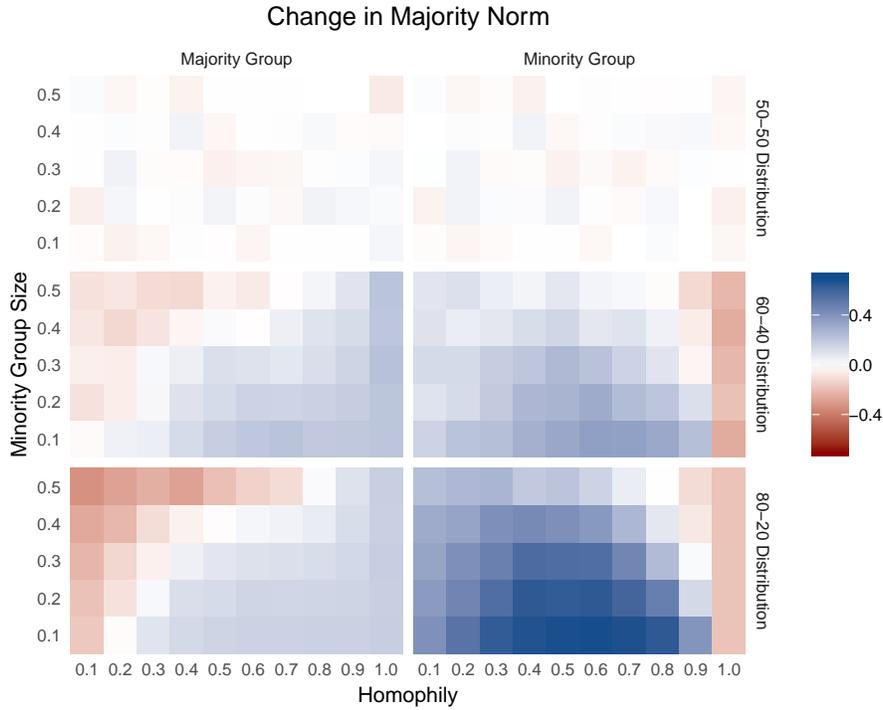}
    \caption{Change in Majority Norm for majority and minority group. This set of heatmaps displays the influence of network homophily/heterophily, group size, and initial norm distributions on change in the majority norm. Each square represents the degree to which representation of the majority norm has increased or decreased in each group. Darker blue means shift towards the majority norm and darker orange means shift towards the minority norm. When norms are initially distributed equally (50-50, top row), the change in group norm difference is essentially random and does not depend on the properties of network structure and group size. When norms are initially distributed unequally e.g. (80-20, bottom row) we observe the impact of homophily and group size. For small homophily values, majority members are more likely to change their norm to the minority norm. As homophily increases, the majority and the minority are both likely to adopt the majority norm (until $h$ = 1, when the pattern is reversed). In general, as the minority group increases in size, it is more likely to retain its own norm and influence the majority.}
    \label{fig:ChangeInProportionMajorityNorm}
\end{figure}

Our first interest is how the representation of norms within groups changes, using our Change in Majority Norm metric (see section~\ref{sec:outcome_metrics}). Figure \ref{fig:ChangeInProportionMajorityNorm} displays the results of the simulations, showing how this is influenced by homophily/heterophily, group sizes, and initial group norm distributions. 

This visualization highlights several findings. First, the effect of homophily and group size on the results is clearest when the initial group norm distribution is 80-20. That is, when norms are highly aligned with group membership, the influence of network structure is most pronounced. When the initial norm distributions are 50-50 in each group, the change in norm proportion is random - the system is not changing systematically even with varying levels of group sizes and homophily/heterophily. Second, the pattern of results for majority and minority groups are distinct. In the majority group, high heterophily (i.e. a greater proportion of connections to the minority) leads to stronger adoption of the minority norm. Similarly, as the size of the minority group increases, the majority group is more likely to adopt the minority group norm. Within Granovetter's threshold model, this is very reasonable: increased minority group size makes it more likely for a majority group member to be connected to members of the minority group and take on the minority norm.

The minority group adopts the majority norm most when homophily is middling and the minority group is small. The minority group maintains or increases its own norm most when it is relatively large, or when homophily is very high or very low. This suggests the operation of multiple mechanisms at different intersections of homophily and group size (see analytical derivations in the appendix). When the network is highly homophilic, the minority maintains its own norm because it is selectively attached to members of its own group, thereby avoiding exposure to majority-group influence. When the network is heterophilic and the minority group is small, the minority is also more able to maintain its own norm. This is because this network parameterization results in minority group members becoming hubs: each majority group member connects to minority group members, and there are not many of them. This means each minority group agent has disproportionate influence. With a large minority group, minority agents have a higher likelihood to be attached to other minority agents, again making it more likely that they will maintain their own norm distribution. The results of the simulation are in agreement with our analytical results provided in the appendix.

\subsection{Change in Group Norm Difference}

\begin{figure}
    \centering
    \includegraphics[width=12cm]{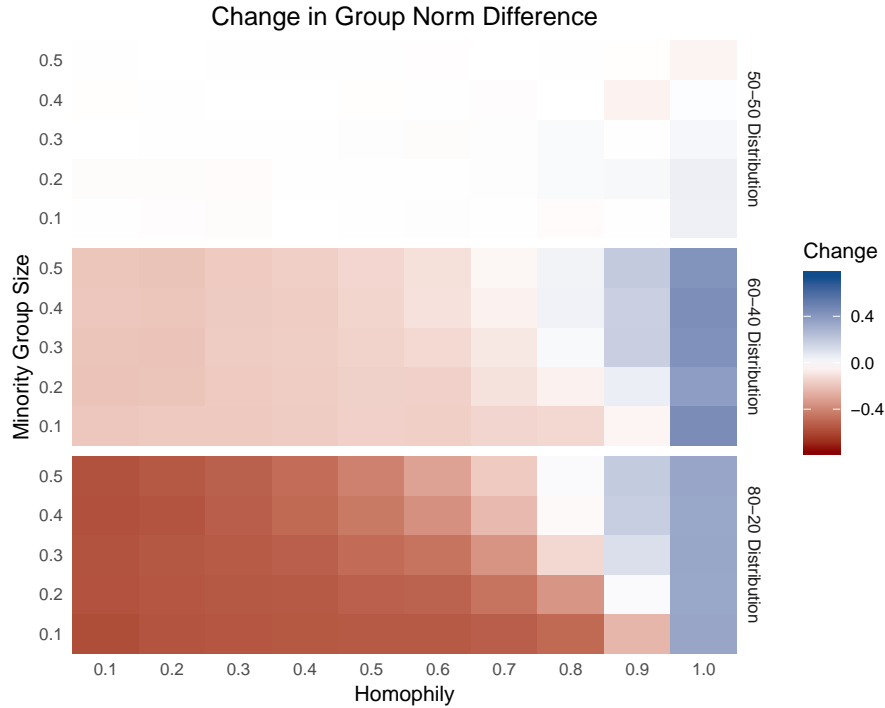}
    \caption{Change in Group Norm Difference. The more negative this number, the more similar the groups have become in their norm distributions; the more positive, the more the groups have diverged in their norm distributions.  In the 50-50 distribution condition, we see that there is no systematic effect of the two groups becoming more normatively similar. In the 80-20 distribution, the  network structure results in strong mutual conformity unless homophily and/or minority group size are very high.}
    \label{fig:DifferenceReduction}
\end{figure}

Our second point of interest is the degree to which the two groups become more similar in their group norm distributions. To address this, we use our Change in Group Norm Difference metric (see section~\ref{sec:outcome_metrics}). The more negative this number, the more similar the groups have become in their norm distributions; the more positive, the more the groups have diverged in their norm distributions. Figure \ref{fig:DifferenceReduction} displays the results of the simulations.

As with the change in norm proportions, the effects of homophily and group proportion are clearest when the initial group norm distribution is strongly associated with group membership (i.e. 80-20 initial group norm distribution). In this case, we can see there is a strong pattern of the two groups moving towards similar norm distributions (i.e. reduce their differences). This pattern is less pronounced or reversed as homophily increases and minority group size increases. This suggests that heterophily is important for producing between-group norm similarity, while high homophily may actually increase between-group norm difference. 

\subsection{Conflict Ties}
\begin{figure}
    \centering
    \includegraphics[width=12cm]{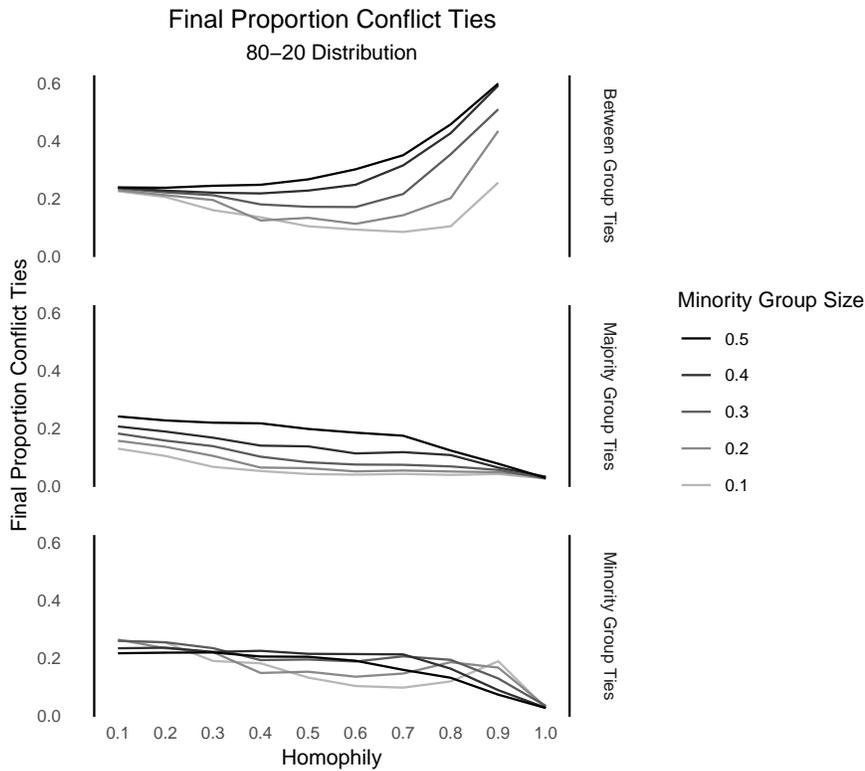}
    \caption{Final Proportion of Conflict Ties in 80-20 Initial Norm Distribution. We see that highly homophilic networks still have relatively high potential for between-group conflict (top row). In contrast, when there is low homophily, the between-group conflict decreases. A reverse pattern appears for within-group conflict ties (second and third rows). As homophily increases within-group conflict decreases.}
    \label{fig:ConflictAtEnd}
\end{figure}

Our third question revolves around the remaining potential for normative conflict, once the simulation has run. For this, we look at the proportion of within- and between-group ties that are Conflict Ties at the end of the simulation. Figure \ref{fig:ConflictAtEnd} shows the results of our simulation for proportion of within-group and between-group ties that are conflict ties. We display results from the 80-20 initial group norm distribution, where group membership and initial norm distribution are closely connected. As with the prior analyses, the results of the 50-50 initial norm distribution were essentially random, and the pattern in the 60-40 initial norm distribution is similar to the 80-20 case but not as strong. 

Comparing the three graphs in Figure \ref{fig:ConflictAtEnd}, we see that the level of network homophily determines the trade-off between intergroup and intragroup conflict. In high-homophily networks high potential for intergroup conflict remains at the end of the simulation, but there is little potential for intragroup conflict. In contrast, high-heterophily networks have very little remaining potential for intergroup conflict, but slightly higher potential for intragroup conflict. 

The role of minority group size also emerges clearly in Figure \ref{fig:ConflictAtEnd}. For between-group ties and majority-group ties, having a small minority group reduces potential conflict. This effect is relatively consistent across all the levels of homophily, though it is more exaggerated at more extreme ones. Within the minority group, group size does not appear to have as consistent of an effect on conflict ties. 

\section{Discussion and Conclusion}

We see three important strands in our pattern of results. First, they speak to the degree to which the alignment of initial group norm distributions and group membership is crucial for the process of reaching normative consensus. Second, they point towards the impact of homophily and heterophily in balancing between ingroup and outgroup conflict. Finally, they point towards strategies that could be used to maintain minority norms in minority groups and to avoid large-scale assimilation. 

\subsection{The Alignment of Norms and Group Membership}
One clear result of our simulation is that, in a system with conflicting norms, substantive change occurs only when the norm is highly aligned with group membership. In our model, this took the form of an 80-20 initial norm distribution, where 80\% of the majority group but only 20\% of the minority group initially held the majority norm. In cases where the norm was not aligned with group membership (50-50 initial norm distribution, top row of Figure \ref{fig:ChangeInProportionMajorityNorm} and Figure \ref{fig:DifferenceReduction}), we do not observe any clear globally dominant norm at the end of the simulation. Even in cases with a relatively large majority group (minority group only 10\% of the network), there was no particular norm change because the social influence of the majority group was evenly split between two norms. When the norm is moderately aligned with group membership (60-40 initial norm distribution), we see intermediate results - not entirely random as with the 50-50, but less clear than when the norm is strongly associated with group membership. 

In intergroup situations, we see that group-level and system-level influence arises not out of small pockets of extremely strong beliefs (i.e. the small minority group in an 80-20 initial norm distribution), but rather out of the consistent homogeneous norm of a majority group. There are cases of normative disagreement that take on proportions like this - our headscarf example from the beginning, for instance, showed 81\% of Germans in favor of banning the headscarf in public institutions, with only 15\% contradicting that opinion. Though such distinct norms are likely to be newsworthy, perhaps there are many instances of intergroup norm non-conflict that receive less attention. Newspapers are unlikely to report that two neighbors from different cultural backgrounds both like to eat dinner with their families, but it may be important for collective cohesion nonetheless. 

This also supports prior literature suggesting that groups with consensual norms are most likely to prompt normative change in their outgroup. A recent survey in the U.S., for instance, indicated a 50-50 split on whether football players should be required to stand during the national anthem \cite{FootballPlayersPoll}. In this case, Americans as a single majority group are unlikely to exert much normative force on out-group members (e.g. Canadians) about this issue. If we consider subgroups of Americans (i.e. Republicans and Democrats), this norm may be much more strongly associated with group membership and thus more likely to have an effect. 

Our model focuses on the shift in a specific norm within a network. This fits our interest in descriptive norms, though real cultural practices might be whole clusters of normative behaviors rather than single binary norms. A contrast between Jewish and Christian people, for example, is not only that they attend different religious services, but also that they can have distinct injunctive norms around weekend hours, food, and marriage that are culturally transmitted. One option is to consider the norm in our model as an aggregate, i.e. not a single behavior, but a cluster of group-based behaviors. Another option is to consider the norm in our model to be the behavior which people notice within a specific context.

\subsection{Homophily Balances In-Group and Between-Group Conflict}

One of the primary aims of this model was to understand when and how subgroups would conform to each other. In Figure \ref{fig:DifferenceReduction}, we see that between-group differences in norms are clearly reduced by the norm updating process, particularly when norms are strongly associated with group membership (i.e. 80-20 initial norm distribution). Except in cases of large minority groups or extremely homophilic networks, there is a meaningful reduction in between-group norm differences: the groups become more similar as the individual agents change their norms. Looking at Figure \ref{fig:ChangeInProportionMajorityNorm}, it is clear that most of the norm change happens in the minority group - they tend to update their norm to that of the majority group, especially when homophily is intermediate and the minority group not large. In contrast, we see the majority group leaving their norm and adopting the minority norm when the network is extremely heterophilic (i.e. $h = 0.1$) (Figure \ref{fig:ChangeInProportionMajorityNorm}). This occurs while the minority group is updating to the majority norm. In this situation, heterophily is so strong that the members of the majority group are disproportionately exposed to the norm of the minority group; this allows for strong influence of the minority group even when the minority is quite small. Thus, though the system overall produces mutual conformity, the level of homophily balances which group is changing their norms to accommodate to the other group. 

In Figure \ref{fig:ConflictAtEnd}, homophily again balances group-level and system-level outcomes when considering the remaining potential for conflict within the network. When the network is heterophilic or neutral, few between-group conflict ties remain. In contrast, when the network is very homophilic, we see the potential for intergroup conflict almost doubled. The reverse is true for within-group conflict ties. When the network is heterophilic or neutral, a fair number of within-group conflict ties remain. When the network is very homophilic, this potential intragroup conflict is reduced by at least half. Thus, we see that both in terms of which group changes their norms and the potential conflict that remains, homophily balances between group-level and system-level outcomes. 

\subsection{Strategies to Maintain Minority Norms}

The maintenance of a cultural identity, partially defined by normative practice, can be extremely important. Our simulation lends support to three methods for maintaining minority cultural practice visibly employed by minority groups in reality: isolationism, adopting positions of influence, and increasing the group size of one's minority. Within the model, the minority group was best able to maintain their own norm in extremely homophilic networks, extremely heterophilic networks, and when their group was large.

Extremely homophilic networks in our simulation mimic strongly isolationist cultures in reality. Such isolation can be imposed upon a minority group (e.g. being excluded from mainstream culture), but can also be sought out as a source of cultural affirmation and strength (e.g. resisting assimilation into mainstream culture) \cite{berry2005acculturation}. This latter motivation has been expressed by groups as different as the Amish in the United States and anti-capitalist leadership in China. The recognition of community-level benefits of culturally affirming and relatively homogeneous environments can be seen in the push to maintain historically black colleges and universities, even as black students in America have increasing access to other institutions \cite{HBCUs}. Though isolationism may draw critique as backward-looking, it can be a deep recognition that intergroup contact can fundamentally affect the culture of a minority group.

Extremely heterophilic networks in our simulation, in contrast, are closely related to minority groups which attempt to have their members in positions of overall societal power. Rather than completely preserving group norms through isolation, this strategy attempts to change the larger culture by exerting influence on the majority. This can be observed in efforts to get members of minority groups elected to positions of power, with the explicit goal of increasing minority voice in the government. By holding positions of power within a larger society, minority group members can become hubs to spread their own group norms. 

The final strategy we can relate to our results is to increase one's group size. The logic here is fairly straightforward: the larger a group, the greater chance it has of influencing the whole system. We can see this strategy in the tendency of minority group members to define their groups expansively, stressing the similarities with the majority group \cite{wimmer2013ethnic}, and the converse tendency of majority group members to define their groups strictly \cite{DovidioIngroupOugroup}. 

The three strategies which emerge from our study are far from a complete set; there are many other strategies well outside the scope of our current work. For example, minority groups actively resist norm change \cite{xie2011social}, cultural institutions formally negotiate over cultural practices, and younger generations modify their inherited cultural practices. We leave model-based exploration of these possibilities for future work. 

\subsection{Limitations and Future Directions}  

In the effort to construct a parsimonious model from existing theory, we acknowledge that there are many assumptions in this model that could be productively expanded. First, one could incorporate more than two groups, or multiple kinds of interpersonal ties. Second, one could make the model more realistic by having a series of inter-correlated norms held by each group, such that individuals have different thresholds to specific norms, or a different weight for norms depending on in-group membership of neighbors. Third, one could integrate psychological theories of preferential information processing to have agents differentially weight the norms expressed by their neighbors based on shared in-group membership. Such modifications would allow us to expand from descriptive to injunctive norms, involving higher-order cognitive processes such as persuasion \cite{cialdini2004social} and contrast with personal values \cite{wei2016moderating} that could be modelled in agents. Finally, it would be valuable to explore other distributions of thresholds within the network to explore more realistic and complex scenarios. These further developments  would also increase the options for validating this model against real world data (e.g. gathering experimental data or found social network data measuring intergroup norm spread). Thus, continuing to grow this work can increase its contribution to the nexus between networks, social norms and conflict.

Despite these limitations, the current study provides a novel and meaningful insight by providing a streamlined example of how group size and homophily can affect the adoption and maintenance of group-affiliated norms. We have shown that even in this simplified version of reality, differences in group proportions and homophily have different effects for majority and minority groups, and can affect the degree to which groups eventually adopt similar distributions of norms. We also contribute to the exciting interdisciplinary growth of computational social science by providing a novel agent-based model that includes both structure of social networks and social influence in one framework.

Finally, we hope that this work contributes to existing knowledge on assimilation, acculturation, and between-group conflict over norms. Our simulation demonstrates that assimilation is most likely at low (heterophilic networks) and intermediate levels of homophily. At intermediate levels, the minority group largely conforms to the majority group. This moves the system towards collective harmony, but does so at the cost of the minority group giving up its own norms. At low levels of homophily, when minority group members have a structural advantage within the network (i.e. central positions with many ties), we see accommodation from both directions: the minority members take on the norm of the majority group, but the majority members also take on the norm of the minority group. Taken together, these suggest that collective harmony is maximized when groups are interconnected, and that this is accompanied by the dispersion of minority norms when there is a strong preference for out-group contact. 

\begin{acknowledgement}
The authors thank the organizers and participants of the BIGSSS summer school for fruitful discussions and feedback, James A. Kitts and the Social Network Group at the Sociology Department at the University of Groningen for insightful comments and feedback, and the Regional Computing Center of the University of Cologne (RRZK) for providing computing time on the DFG-funded High Performance Computing (HPC) system CHEOPS as well as support. Rocco Paolillo completed this work while on the programme EU COFUND BIGSSS-departs, Marie Skłodowska-Curie grant agreement No. 713639. Natalie Gallagher completed this work while on the Graduate Research Fellowship from the National Science Foundation.
\end{acknowledgement}

\begin{contribution}
All authors jointly came up with the idea and research questions. J.K. implemented the simulation model, wrote the method part and contributed to the theoretical background part. N.G. analyzed results and wrote results and discussion. Z.M.K, R.P. and L.P. conducted literature research and wrote the theoretical background part. F.K. provided mentoring, computed analytical derivations and contributed to results and discussion.
\end{contribution}

\section*{Appendix: Analytical Derivations for Norm Endorsement}
\label{sec:analytical_derivations}
\addcontentsline{toc}{section}{Appendix}

In this appendix, we derive the probabilities of norm endorsement in each group using the mean-field approach. This analysis enables us to gain insights on the relationship between the model parameters of homophily, group size, and group norm distribution. In addition, the analytical derivations help us to interpret the outcome of the simulations in section \ref{sec:simulation_results}. 

More specifically, we calculate the probabilities of a minority agent to update to the majority norm and \textit{vice versa}. We use mean-field approximation (also known as the deterministic approximation) which means that we look at the average behavior of the group in an equilibrium state \cite{marro2005nonequilibrium}. That means, we do not consider the changes over time and the heterogeneity of the agents. Nevertheless, the mean-field approach gives us a useful insight on forecasting the overall behavior of the system. Let us assume that the minority is denoted by $a$ and the majority is denoted by $b$. Two norms are denoted by norm $A$ and norm $B$. Homophily is denoted by $h$ and group proportion is denoted by $g$. In order to calculate the probability of a minority agent to update the majority norm ($B$), we need to estimate the probability of a minority to be connected to majority agents ($p_{ab}$) and the probability of the minority agent to be connected to minority agents ($p_{aa}$). Since our agent-based model assumes a preferential attachment mechanism and defines group proportion ($g$), the probability of two agents to be connected depends on their homophily ($h$) and the degree of the agent ($k$). Link formation is a combination of two mechanisms, namely homophily and preferential attachment, and thus the probability of connectivity follows a non-linear function. To estimate the link probabilities, apart from homophily, we need to estimate the degree growth function ($C$) of each group of agents. The degree growth determines the attractivity of the agents with regards to their degree. The degree growth in this model follows a polynomial function of order three with one valid solution and it can be calculated numerically \cite{karimi2018homophily}:

\begin{equation}
   C = \left(g\left(1 + \dfrac{hC}{hC + (1-h)(2 - C)}\right) + (1-g)\dfrac{(1-h)C}{h(2 - C) + (1-h)C}\right)\;.
\end{equation}

The probability of two agents of group $a$ ($p_{aa}$) and two agents of group $b$ ($p_{bb}$) to be connected are:
\begin{linenomath*}
\begin{equation}
\label{eq:paa}
\begin{split}
p_{aa} &= \frac{hC}{hC+(1-h)(2-C)}\;,\\
p_{bb} &= \frac{h(2-C)}{h(2-C)+(1-h)C}\;.
\end{split}
\end{equation}
\end{linenomath*}

In addition, the degree growth function has the following relation to the probability of linkage \cite{karimi2018homophily}:

\begin{linenomath*}
\begin{equation}
    C = g(1 + p_{aa}) + (1-g) p_{ba}\;.
\label{eq:C}
\end{equation}
\end{linenomath*}

The probability of a minority agent to update to the majority norm ($f_{aB}$) depends on the probability of being connected to majority ($p_{ab}$) and minority ($p_{aa}$). Thus, for a minority agent, the fraction of neighbors with norm $B$ is:
\begin{equation}
\label{eq:f_B1}
    f_{aB} = \frac{ p_{aa}p_{aB} + p_{ab}p_{bB}}{p_{aa} + p_{ab}}\;.
\end{equation}
The numerator consists of two parts; the probability of connecting to another minority with norm $B$ ($p_{aa}p_{aB}$) and the probability of connecting to majority with norm $B$ ($p_{ab}p_{bB}$). To estimate the fraction, the nominator should be divided by the total probability of connectivity between the majority to majority and minority. Inserting Eq.(\ref{eq:paa}) into Eq.(\ref{eq:f_B1}), we find:

\begin{equation}
\label{eq:f_B2} 
    f_{aB} = \frac{ (\frac{hC}{hC+(1-h)(2-C)}) (p_{aB}-p_{bB}) + (2-\frac{h(2-C)}{h(2-C)+(1-h)C})p_{bB}}{2 -(\frac{h(2-C)}{h(2-C)+(1-h)C}) }\;.
\end{equation}

Similar relation can be found for the probability of a majority agent to update to the minority norm ($f_{bA}$):
\begin{equation}
\label{eq:f_A1}
    f_{bA} = \frac{ p_{bb}p_{bA} + p_{ba}p_{aA}}{p_{bb} + p_{ba}}\;.
\end{equation}

\begin{figure}[H]
\centering
\includegraphics[width=12cm]{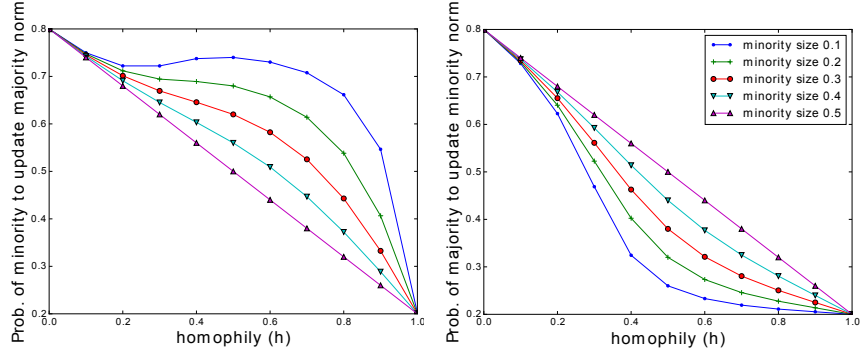}
\caption{Analytical results for the probability of minority (left) and majority (right) to update to the norm of the other group. Initial norm proportion is set to 20-80. We observe asymmetrical results as the group balance deviates from 50-50 condition. For small values of homophily $0 \leq h \leq 0.2$ we observe similar behavior for majority and minority. However, as homophily increases,  we observe that minority members update their norm to that of the majority with high probability while majority does not update to the minority norm. The asymmetric relation is more pronounced as the minority group size decreases.} 
\label{fig:analytic}
\end{figure}

Figure \ref{fig:analytic} displays the analytical results derived from the above derivations. It is interesting to note that the update to the norm of other group follows a nonlinear and asymmetrical trend both for the minority and the majority. In the intermediate level of homophily ($0.5<h<0.8$), while the majority members resist to switch its norm to minority norm, the minority updates to the majority norm with high probability. That would create a higher advantage for the majority norm to persist and stabilize. Only when homophily is very high, the probability of the minority members to update to the majority norm starts decreasing. As the minority size shrinks, the inequality in norm adoption increases.  

\bibliography{ref} 

\begin{thebibliography}{79}
\providecommand{\natexlab}[1]{#1}
\providecommand{\url}[1]{\texttt{#1}}
\expandafter\ifx\csname urlstyle\endcsname\relax
  \providecommand{\doi}[1]{doi: #1}\else
  \providecommand{\doi}{doi: \begingroup \urlstyle{rm}\Url}\fi

\bibitem[Granovetter(1978)]{granovetter1978threshold}
Mark Granovetter.
\newblock Threshold models of collective behavior.
\newblock \emph{American Journal of Sociology}, 83\penalty0 (6):\penalty0
  1420--1443, 1978.

\bibitem[Bicchieri and Mercier(2014)]{bicchieri2014norms}
Cristina Bicchieri and Hugo Mercier.
\newblock \emph{Norms and beliefs: How change occurs}, pages 37--54.
\newblock Springer, New York, 2014.

\bibitem[Forsyth(2018)]{forsyth2018group}
Donelson~R Forsyth.
\newblock \emph{Group dynamics}.
\newblock Wadsworth Cengage Learning, Belmont, 2018.

\bibitem[Bicchieri(2006)]{bicchieri2005grammar}
Cristina Bicchieri.
\newblock \emph{The grammar of society: The nature and dynamics of social
  norms}.
\newblock Cambridge University Press, Cambridge, 2006.

\bibitem[House(2018)]{house2018social}
Bailey~R House.
\newblock How do social norms influence prosocial development?
\newblock \emph{Current Opinion in Psychology}, 20:\penalty0 87--91, 2018.

\bibitem[Melnyk et~al.(2010)Melnyk, van Herpen, and van
  Trijp]{melnyk2010influence}
Vladimir Melnyk, Erica van Herpen, and Hans C.~M. van Trijp.
\newblock The influence of social norms in consumer decision making: {a}
  meta-analysis.
\newblock \emph{Advances in Consumer Research}, 37\penalty0 (1):\penalty0
  463--464, 2010.

\bibitem[Cialdini et~al.(1990)Cialdini, Reno, and Kallgren]{cialdini1990focus}
Robert~B Cialdini, Raymond~R Reno, and Carl~A Kallgren.
\newblock A focus theory of normative conduct: recycling the concept of norms
  to reduce littering in public places.
\newblock \emph{Journal of Personality and Social Psychology}, 58\penalty0
  (6):\penalty0 1015--1026, 1990.

\bibitem[Megens and Weerman(2010)]{megens2010attitudes}
Kim C. I.~M. Megens and Frank~M Weerman.
\newblock Attitudes, delinquency and peers: {t}he role of social norms in
  attitude-behaviour inconsistency.
\newblock \emph{European Journal of Criminology}, 7\penalty0 (4):\penalty0
  299--316, 2010.

\bibitem[Cialdini(2007)]{cialdini2007influence}
Robert~B Cialdini.
\newblock \emph{Influence: The psychology of persuasion}.
\newblock HarperCollins Publishers, New York, 2007.

\bibitem[Fehr and G{\"a}chter(2000)]{fehr2000cooperation}
Ernst Fehr and Simon G{\"a}chter.
\newblock Cooperation and punishment in public goods experiments.
\newblock \emph{The American Economic Review}, 90\penalty0 (4):\penalty0
  980--994, 2000.

\bibitem[Oraby et~al.(2014)Oraby, Thampi, and Bauch]{OrabyThampiBauch2014}
Tamer Oraby, Vivek Thampi, and Chris~T. Bauch.
\newblock The influence of social norms on the dynamics of vaccination
  behaviour for peaedriatic infectious diseases.
\newblock \emph{Proceedings of The Royal Society B: Biological Sciences},
  281\penalty0 (1780):\penalty0 1--8, 2014.

\bibitem[Haines and Spear(1996)]{haines1996changing}
Michael Haines and Sherilynn~F Spear.
\newblock Changing the perception of the norm: {a} strategy to decrease binge
  drinking among college students.
\newblock \emph{Journal of American College Health}, 45\penalty0 (3):\penalty0
  134--140, 1996.

\bibitem[Hogg and Reid(2006)]{hogg2006social}
Michael~A Hogg and Scott~A Reid.
\newblock Social identity, self-categorization, and the communication of group
  norms.
\newblock \emph{Communication Theory}, 16\penalty0 (1):\penalty0 7--30, 2006.

\bibitem[Wimmer(2013)]{wimmer2013ethnic}
Andreas Wimmer.
\newblock \emph{Ethnic boundary making: {I}nstitutions, power, networks}.
\newblock Oxford University Press, New York, 2013.

\bibitem[K{\i}l{\i}{\c{c}} et~al.(2008)K{\i}l{\i}{\c{c}}, Saharso, and
  Sauer]{kilicc2008introduction}
Sevgi K{\i}l{\i}{\c{c}}, Sawitri Saharso, and Birgit Sauer.
\newblock Introduction: {t}he veil: {d}ebating citizenship, gender and
  religious diversity.
\newblock \emph{Social Politics: International Studies in Gender, State \&
  Society}, 15\penalty0 (4):\penalty0 397--410, 2008.

\bibitem[{Zeit Online}(2019)]{zeit_burqa}
{Zeit Online}.
\newblock
  \url{https://www.zeit.de/politik/ausland/2018-10/frankreich-un-menschenrechtsausschuss-burka-verbot-vollverschleierung-menschenrechte-religionsfreiheit},
  2019.
\newblock Accessed 16 May 2019.

\bibitem[McKenzie(2019)]{cnn_veil}
Sheena McKenzie.
\newblock Germany could impose partial ban on face veils, officials say.
\newblock
  \url{https://edition.cnn.com/2016/08/19/europe/germany-veil-ban/index.html},
  2019.
\newblock Accessed 16 May 2019.

\bibitem[{Infratest Dimap}(2018)]{dimap_burqa}
{Infratest Dimap}.
\newblock
  \url{https://www.infratest-dimap.de/umfragen-analysen/bundesweit/umfrage/aktuell/grosse-mehrheit-der-deutschen-plaediert-fuer-burka-verbot/},
  2018.
\newblock Accessed 01 Oct 2018.

\bibitem[Kleck(1996)]{kleck1996crime}
Gary Kleck.
\newblock Crime, culture conflict and the sources of support for gun control:
  {a} multilevel application of the general social surveys.
\newblock \emph{American Behavioral Scientist}, 39\penalty0 (4):\penalty0
  387--404, 1996.

\bibitem[Celinska(2007)]{celinska2007individualism}
Katarzyna Celinska.
\newblock Individualism and collectivism in america: The case of gun ownership
  and attitudes toward gun control.
\newblock \emph{Sociological Perspectives}, 50\penalty0 (2):\penalty0 229--247,
  2007.

\bibitem[Marecek et~al.(2017)Marecek, Macleod, and
  Hoggart]{marecek2017abortion}
Jeanne Marecek, Catriona Macleod, and Lesley Hoggart.
\newblock Abortion in legal, social, and healthcare contexts.
\newblock \emph{Feminism \& Psychology}, 27\penalty0 (1):\penalty0 4--14, 2017.

\bibitem[Evans(2002)]{evans2002polarization}
John~H Evans.
\newblock Polarization in abortion attitudes in {U.S.} religious traditions,
  1972-1998.
\newblock \emph{Sociological Forum}, 17\penalty0 (3):\penalty0 397--422, 2002.

\bibitem[Rodenberg and Wagenaar(2016)]{rodenberg2016essentializing}
Jeroen Rodenberg and Pieter Wagenaar.
\newblock Essentializing {"Black Pete"}: {c}ompeting narratives surrounding the
  {S}interklaas tradition in the {N}etherlands.
\newblock \emph{International Journal of Heritage Studies}, 22\penalty0
  (9):\penalty0 716--728, 2016.

\bibitem[Simmel(2009)]{simmel2009sociology}
Geog Simmel.
\newblock \emph{Sociology: {i}nquiries into the constructions of social forms.
  {V}olume {I}}.
\newblock Brill, Leiden, 2009.
\newblock {Translated and edited by Anthony J. Blasi, Anton K. Jacobs and
  Mathew Kanjirathinkal}.

\bibitem[Tuckman(1965)]{tuckman1965developmental}
Bruce~W Tuckman.
\newblock Developmental sequence in small groups.
\newblock \emph{Psychological Bulletin}, 63\penalty0 (6):\penalty0 384--399,
  1965.

\bibitem[Packer and Miners(2014)]{packer2014tough}
Dominic~J Packer and Christopher~T.H. Miners.
\newblock Tough love: {t}he normative conflict model and a goal system approach
  to dissent decisions.
\newblock \emph{Social and Personality Psychology Compass}, 8\penalty0
  (7):\penalty0 354--373, 2014.

\bibitem[Arapoglou(2012)]{arapoglou2012diversity}
Vassilis~P Arapoglou.
\newblock Diversity, inequality and urban change.
\newblock \emph{European Urban and Regional Studies}, 19\penalty0 (3):\penalty0
  223--237, 2012.

\bibitem[Crul(2016)]{crul2016super}
Maurice Crul.
\newblock Super-diversity vs. assimilation: {h}ow complex diversity in
  majority-minority cities challenges the assumptions of assimilation.
\newblock \emph{Journal of Ethnic and Migration Studies}, 42\penalty0
  (1):\penalty0 54--68, 2016.

\bibitem[Fiorina and Abrams(2008)]{fiorina2008political}
Morris~P Fiorina and Samuel~J Abrams.
\newblock Political polarization in the {A}merican public.
\newblock \emph{Annual Review of Political Science}, 11\penalty0 (1):\penalty0
  563--588, 2008.

\bibitem[Neumann(2008)]{neumann2008homo}
Martin Neumann.
\newblock Homo socionicus: {a} case study of simulation models of norms.
\newblock \emph{Journal of Artificial Societies and Social Simulation},
  11\penalty0 (4):\penalty0 6, 2008.

\bibitem[Latan{\'e}(1981)]{latane1981psychology}
Bibb Latan{\'e}.
\newblock The psychology of social impact.
\newblock \emph{American Psychologist}, 36\penalty0 (4):\penalty0 343--356,
  1981.

\bibitem[Flache et~al.(2017)Flache, M{\"a}s, Feliciani, Chattoe-Brown,
  Deffuant, Huet, and Lorenz]{flache2017models}
Andreas Flache, Michael M{\"a}s, Thomas Feliciani, Edmund Chattoe-Brown,
  Guillaume Deffuant, Sylvie Huet, and Jan Lorenz.
\newblock Models of social influence: {t}owards the next frontiers.
\newblock \emph{Journal of Artificial Societies \& Social Simulation},
  20\penalty0 (4):\penalty0 2, 2017.

\bibitem[Kalesan et~al.(2016)Kalesan, Villarreal, Keyes, and
  Galea]{kalesan2016gun}
Bindu Kalesan, Marcos~D. Villarreal, Katherine~M Keyes, and Sandro Galea.
\newblock Gun ownership and social gun culture.
\newblock \emph{Injury Prevention}, 22\penalty0 (3):\penalty0 216--220, 2016.

\bibitem[Tankard and Paluck(2017)]{TankardPaluck2017}
Margaret~E. Tankard and Elizabeth~L. Paluck.
\newblock The effect of a supreme court decision regarding gay marriage on
  social norms and personal attitudes.
\newblock \emph{Psychological Science}, 28\penalty0 (9):\penalty0 1334--1444,
  2017.

\bibitem[McPherson et~al.(2001)McPherson, Smith-Lovin, and
  Cook]{mcpherson2001birds}
Miller McPherson, Lynn Smith-Lovin, and James~M Cook.
\newblock Birds of a feather: {h}omophily in social networks.
\newblock \emph{Annual Review of Sociology}, 27\penalty0 (1):\penalty0
  415--444, 2001.

\bibitem[Lozares et~al.(2014)Lozares, Verd, Cruz, and
  Barranco]{lozares2014homophily}
Carlos Lozares, Joan~Miquel Verd, Irene Cruz, and Oriol Barranco.
\newblock Homophily and heterophily in personal networks. from mutual
  acquaintance to relationship intensity.
\newblock \emph{Quality \& Quantity}, 48\penalty0 (5):\penalty0 2657--2670,
  2014.

\bibitem[Stehl{\'e} et~al.(2013)Stehl{\'e}, Charbonnier, Picard, Cattuto, and
  Barrat]{stehle2013gender}
Juliette Stehl{\'e}, Fran{\c{c}}ois Charbonnier, Tristan Picard, Ciro Cattuto,
  and Alain Barrat.
\newblock Gender homophily from spatial behavior in a primary school: {a}
  sociometric study.
\newblock \emph{Social Networks}, 35\penalty0 (4):\penalty0 604--613, 2013.

\bibitem[Jadidi et~al.(2017)Jadidi, Karimi, Lietz, and
  Wagner]{jadidi2017gender}
Mohsen Jadidi, Fariba Karimi, Haiko Lietz, and Claudia Wagner.
\newblock Gender disparities in science? dropout, productivity, collaborations
  and success of male and female computer scientists.
\newblock \emph{Advances in Complex Systems}, 21\penalty0 (3-4):\penalty0
  1--23, 2017.

\bibitem[Mislove et~al.(2010)Mislove, Viswanath, Gummadi, and
  Druschel]{mislove2010you}
Alan Mislove, Bimal Viswanath, Krishna~P Gummadi, and Peter Druschel.
\newblock You are who you know: {i}nferring user profiles in online social
  networks.
\newblock In \emph{Proceedings of the third {ACM} international conference on
  Web search and data mining}, pages 251--260. ACM, 2010.

\bibitem[Centola(2010)]{centola2010spread}
Damon Centola.
\newblock The spread of behavior in an online social network experiment.
\newblock \emph{Science}, 329\penalty0 (5996):\penalty0 1194--1197, 2010.

\bibitem[Christakis and Fowler(2008)]{christakis2008collective}
Nicholas~A Christakis and James~H Fowler.
\newblock The collective dynamics of smoking in a large social network.
\newblock \emph{The New England Journal of Medicine}, 358\penalty0
  (21):\penalty0 2249--2258, 2008.

\bibitem[Christakis and Fowler(2007)]{christakis2007spread}
Nicholas~A Christakis and James~H Fowler.
\newblock The spread of obesity in a large social network over 32 years.
\newblock \emph{The New England Journal of Medicine}, 357\penalty0
  (4):\penalty0 370--379, 2007.

\bibitem[Aral and Walker(2011)]{aral2011creating}
Sinan Aral and Dylan Walker.
\newblock Creating social contagion through viral product design: {a}
  randomized trial of peer influence in networks.
\newblock \emph{Management {S}cience}, 57\penalty0 (9):\penalty0 1623--1639,
  2011.

\bibitem[Lewis et~al.(2012)Lewis, Gonzalez, and Kaufman]{lewis2012social}
Kevin Lewis, Marco Gonzalez, and Jason Kaufman.
\newblock Social selection and peer influence in an online social network.
\newblock \emph{Proceedings of the National Academy of Sciences of the United
  States of America}, 109\penalty0 (1):\penalty0 68--72, 2012.

\bibitem[Blau(1977)]{blau1977macrosociological}
Peter~M Blau.
\newblock A macrosociological theory of social structure.
\newblock \emph{American {J}ournal of {S}ociology}, 83\penalty0 (1):\penalty0
  26--54, 1977.

\bibitem[Asch(1951)]{asch1951effects}
Solomon~E Asch.
\newblock \emph{Effects of group pressure upon the modification and distortion
  of judgments}, pages 177--199.
\newblock Carnegie Press, Oxford, 1951.

\bibitem[Horcajo et~al.(2010)Horcajo, Petty, and
  Bri{\~n}ol]{horcajo2010effects}
Javier Horcajo, Richard~E Petty, and Pablo Bri{\~n}ol.
\newblock The effects of majority versus minority source status on persuasion:
  {a} self-validation analysis.
\newblock \emph{Journal of Personality and Social Psychology}, 99\penalty0
  (3):\penalty0 498--512, 2010.

\bibitem[Kundu and Cummins(2013)]{kundu2013morality}
Payel Kundu and Denise~Dellarosa Cummins.
\newblock Morality and conformity: {t}he {A}sch paradigm applied to moral
  decisions.
\newblock \emph{Social Influence}, 8\penalty0 (4):\penalty0 268--279, 2013.

\bibitem[Meyers et~al.(2000)Meyers, Brashers, and Hanner]{meyers2000majority}
Ren{\'e}e~A Meyers, Dale~E Brashers, and Jennifer Hanner.
\newblock Majority-minority influence: {i}dentifying argumentative patterns and
  predicting argument-outcome links.
\newblock \emph{Journal of Communication}, 50\penalty0 (4):\penalty0 3--30,
  2000.

\bibitem[Cohen(2003)]{cohen2003party}
Geoffrey~L Cohen.
\newblock Party over policy: {t}he dominating impact of group influence on
  political beliefs.
\newblock \emph{Journal of Personality and Social Psychology}, 85\penalty0
  (5):\penalty0 808--822, 2003.

\bibitem[Bourhis et~al.(1997)Bourhis, Moise, Perreault, and
  Senecal]{bourhis1997towards}
Richard~Y Bourhis, Lena~Celine Moise, Stephane Perreault, and Sacha Senecal.
\newblock Towards an interactive acculturation model: {a} social psychological
  approach.
\newblock \emph{International Journal of Psychology}, 32\penalty0 (6):\penalty0
  369--386, 1997.

\bibitem[Ward et~al.(2010)Ward, Fox, Wilson, Stuart, and
  Kus]{ward2010contextual}
Colleen Ward, Stephen Fox, Jessie Wilson, Jaimee Stuart, and Larissa Kus.
\newblock Contextual influences on acculturation processes: {t}he roles of
  family, community and society.
\newblock \emph{Psychological Studies}, 55\penalty0 (1):\penalty0 26--34, 2010.

\bibitem[Mugny and Papastamou(1982)]{mugny1982minority}
Gabriel Mugny and Stamos Papastamou.
\newblock Minority influence and psycho-social identity.
\newblock \emph{European Journal of Social Psychology}, 12\penalty0
  (4):\penalty0 379--394, 1982.

\bibitem[Nemeth(1986)]{nemeth1986differential}
Charlan~J Nemeth.
\newblock Differential contributions of majority and minority influence.
\newblock \emph{Psychological Review}, 93\penalty0 (1):\penalty0 23--32, 1986.

\bibitem[Hornsey(2008)]{hornsey2008social}
Matthew~J Hornsey.
\newblock Social identity theory and self-categorization theory: {a} historical
  review.
\newblock \emph{Social and Personality Psychology Compass}, 2\penalty0
  (1):\penalty0 204--222, 2008.

\bibitem[Asch(1956)]{asch1956studies}
Solomon~E Asch.
\newblock Studies of independence and conformity: I. {A} minority of one
  against a unanimous majority.
\newblock \emph{Psychological Monographs: General and Applied}, 70\penalty0
  (9):\penalty0 1--70, 1956.

\bibitem[Marques et~al.(1988)Marques, Yzerbyt, and Leyens]{marques1988black}
Jos{\'e}~M Marques, Vincent~Y Yzerbyt, and Jacques-Philippe Leyens.
\newblock The “black sheep effect”: {e}xtremity of judgments towards
  ingroup members as a function of group identification.
\newblock \emph{European Journal of Social Psychology}, 18\penalty0
  (1):\penalty0 1--16, 1988.

\bibitem[Macy and Willer(2002)]{macy2002factors}
Michael~W Macy and Robert Willer.
\newblock From factors to actors: {c}omputational sociology and agent-based
  modeling.
\newblock \emph{Annual Review of Sociology}, 28\penalty0 (1):\penalty0
  143--166, 2002.

\bibitem[Squazzoni et~al.(2014)Squazzoni, Jager, and
  Edmonds]{squazzoni2014social}
Flaminio Squazzoni, Wander Jager, and Bruce Edmonds.
\newblock Social simulation in the social sciences: {a} brief overview.
\newblock \emph{Social Science Computer Review}, 32\penalty0 (3):\penalty0
  279--294, 2014.

\bibitem[Schelling(1971)]{schelling1971dynamic}
Thomas~C Schelling.
\newblock Dynamic models of segregation.
\newblock \emph{Journal of Mathematical Sociology}, 1\penalty0 (2):\penalty0
  143--186, 1971.

\bibitem[Lorenz(2007)]{lorenz2007continuous}
Jan Lorenz.
\newblock Continuous opinion dynamics under bounded confidence: {a} survey.
\newblock \emph{International Journal of Modern Physics C}, 18\penalty0
  (12):\penalty0 1819--1838, 2007.

\bibitem[Zhang and Vorobeychik(2017)]{zhang2017empirically}
Haifeng Zhang and Yevgeniy Vorobeychik.
\newblock Empirically grounded agent-based models of innovation diffusion: {a}
  critical review.
\newblock \emph{Artificial Intelligence Review}, 52\penalty0 (1):\penalty0
  707--741, 2017.

\bibitem[Watts(2002)]{watts2002simple}
Duncan~J Watts.
\newblock A simple model of global cascades on random networks.
\newblock \emph{Proceedings of the National Academy of Sciences of the United
  States of America}, 99\penalty0 (9):\penalty0 5766--5771, 2002.

\bibitem[Karimi et~al.(2018)Karimi, G{\'e}nois, Wagner, Singer, and
  Strohmaier]{karimi2018homophily}
Fariba Karimi, Mathieu G{\'e}nois, Claudia Wagner, Philipp Singer, and Markus
  Strohmaier.
\newblock Homophily influences ranking of minorities in social networks.
\newblock \emph{Scientific Reports}, 8\penalty0 (1):\penalty0 1--12, 2018.

\bibitem[{R Core Team}(2019)]{Rlang}
{R Core Team}.
\newblock R: {a} language and environment for statistical computing.
\newblock \url{https://www.R-project.org/}, 2019.
\newblock Accessed 27 May 2019.

\bibitem[Kohne(2019)]{SimGit}
J.~Kohne.
\newblock Simulating normative conflict.
\newblock \url{https://github.com/JuKo007/SimulatingNormativeConflict. DOI:
  10.5281/zenodo.3183121}, 2019.
\newblock Accessed 07 Jun 2019.

\bibitem[Barab{\'a}si and R{\'e}ka(1999)]{barabasi1999emergence}
Albert-L{\'a}szl{\'o} Barab{\'a}si and Albert R{\'e}ka.
\newblock Emergence of scaling in random networks.
\newblock \emph{Science}, 286\penalty0 (5439):\penalty0 509--512, 1999.

\bibitem[DiMaggio and Garip(2012)]{dimaggio2012network}
Paul DiMaggio and Filiz Garip.
\newblock Network effects and social inequality.
\newblock \emph{Annual Review of Sociology}, 38\penalty0 (1):\penalty0 93--118,
  2012.

\bibitem[Eberlen et~al.(2017)Eberlen, Scholz, and
  Gagliolo]{erleben2017simulate}
Julia Eberlen, Geeske Scholz, and Matteo Gagliolo.
\newblock Simulate this! {A}n introduction to agent-based models and their
  power to improve your research practice.
\newblock \emph{International Review of Social Psychology}, 30\penalty0
  (1):\penalty0 149--160, 2017.

\bibitem[Railsback and Grimm(2019)]{railsback2019agent}
Steven~F Railsback and Volker Grimm.
\newblock \emph{Agent-based and individual-based modeling: {A} practical
  introduction}.
\newblock Princeton University Press, Princeton, 2019.

\bibitem[Lee et~al.(2015)Lee, Filatova, Ligmann-Zielinska, Hassani-Mahmooei,
  Stonedahl, Lorscheid, Voinov, Polhill, Sun, and Parker]{lee2015complexities}
Ju-Sung Lee, Tatiana Filatova, Arika Ligmann-Zielinska, Behrooz
  Hassani-Mahmooei, Forrest Stonedahl, Iris Lorscheid, Alexey Voinov, J~Gary
  Polhill, Zhanli Sun, and Dawn~C Parker.
\newblock The complexities of agent-based modeling output analysis.
\newblock \emph{Journal of Artificial Societies and Social Simulation},
  18\penalty0 (4):\penalty0 4, 2015.

\bibitem[Vandermaas-Peeler et~al.(2018)Vandermaas-Peeler, Cox, Maxine,
  Fisch-Friedman, Griffin, and Jones]{FootballPlayersPoll}
Alex Vandermaas-Peeler, Daniel Cox, Najle Maxine, Molly Fisch-Friedman, Rob
  Griffin, and Robert~P Jones.
\newblock Partisan polarization dominates {T}rump era: Findings from the 2018
  american values survey.
\newblock Resource document. Public Policy Research Institute.
  \url{https://www.prri.org/research/partisan-polarization-dominates-trump-era-findings-from-the-2018-american-values-survey/},
  2018.
\newblock Accessed 27 May 2019.

\bibitem[Berry(2005)]{berry2005acculturation}
John~W Berry.
\newblock Acculturation: {l}iving successfully in two cultures.
\newblock \emph{International Journal of Intercultural Relations}, 29\penalty0
  (6):\penalty0 697--712, 2005.

\bibitem[Franke and DeAngelo(2018)]{HBCUs}
Ray Franke and Linda DeAngelo.
\newblock Degree attainment for black students at {HBCUs} and pwis: {a}
  propensity score matching approach.
\newblock \emph{Paper Presented at the Annual Meeting of the American
  Educational Research Association}, 2018.

\bibitem[Dovidio et~al.(2007)Dovidio, Gaertner, and
  Saguy]{DovidioIngroupOugroup}
J.F. Dovidio, S.L. Gaertner, and T.~Saguy.
\newblock Another view of "we": {m}ajority and minority group perspectives on a
  common ingroup identity.
\newblock \emph{European Review of Social Psychology}, 18\penalty0
  (1):\penalty0 296--330, 2007.

\bibitem[Xie et~al.(2011)Xie, Sreenivasan, Korniss, Zhang, Lim, and
  Szymanski]{xie2011social}
Jierui Xie, Sameet Sreenivasan, Gyorgy Korniss, Weituo Zhang, Chjan Lim, and
  Boleslaw~K Szymanski.
\newblock Social consensus through the influence of committed minorities.
\newblock \emph{Physical Review E}, 84\penalty0 (1):\penalty0 1--8, 2011.

\bibitem[Cialdini and Goldstein(2004)]{cialdini2004social}
Robert~B Cialdini and Noah~J Goldstein.
\newblock Social influence: {c}ompliance and conformity.
\newblock \emph{Annual Review of Psychology}, 55:\penalty0 591--621, 2004.

\bibitem[Wei et~al.(2016)Wei, Zhao, and Zheng]{wei2016moderating}
Zhenyu Wei, Zhiying Zhao, and Yong Zheng.
\newblock Moderating effects of social value orientation on the effect of
  social influence in prosocial decisions.
\newblock \emph{Frontiers in Psychology}, 7:\penalty0 1--9, 2016.

\bibitem[Marro and Dickman(2005)]{marro2005nonequilibrium}
Joaqu{\'\i}n Marro and Ronald Dickman.
\newblock \emph{Nonequilibrium phase transitions in lattice models}.
\newblock Cambridge University Press, Cambridge, 2005.

\end{thebibliography}

\end{document}